\input harvmac
%
\def\eg{{\it e.g.}}
\def\ie{{\it i.e.}}
\noblackbox 

\newcount\figno 
\figno=0 
\def\fig#1#2#3{ 
\par\begingroup\parindent=0pt\leftskip=1cm\rightskip=1cm\parindent=0pt 
\baselineskip=11pt 
\global\advance\figno by 1 
\midinsert 
\epsfxsize=#3 
\centerline{\epsfbox{#2}} 
\vskip 12pt 
{\bf Fig.\ \the\figno: } #1\par 
\endinsert\endgroup\par 
} 
\def\figlabel#1{\xdef#1{\the\figno}} 
\def\encadremath#1{\vbox{\hrule\hbox{\vrule\kern8pt\vbox{\kern8pt 
\hbox{$\displaystyle #1$}\kern8pt} 
\kern8pt\vrule}\hrule}} 
 
\def\frac#1#2{{#1 \over #2}}

\def\semi{\subset\kern-1em\times\;} 

\def\IR{\relax{\rm I\kern-.18em R}}
\def\ZZ{\hbox{\rm Z\kern-.4emZ}}

\Title{\vbox{\baselineskip12pt 
\hbox{hep-th/0009244} 
\hbox{EFI-2000-36} 
\vskip-.5in}} 
{\vbox{\centerline{Partition Sums and Entropy Bounds}  
\bigskip
\centerline{in Weakly Coupled CFT} 
\bigskip}
}
\medskip\bigskip 
\centerline{David Kutasov and Finn Larsen} 
\bigskip\medskip 
\centerline{\it Enrico Fermi Institute and Department of Physics} 
\centerline{\it  University of Chicago,  
Chicago, IL 60637, USA} 
\baselineskip18pt 
\medskip\bigskip\medskip\bigskip\medskip 
\baselineskip16pt 
We use the partition functions on $S^1\times S^n$
of various conformal field theories  in four and six
dimensions in the limit of vanishing coupling to study the 
high temperature thermodynamics. Certain modular properties 
exhibited by the partition functions help to determine the 
finite volume corrections, which play a role in the discussion of 
entropy bounds.

\Date{9/00} 
\lref\adsrev{O.~Aharony, S.~S.~Gubser, J.~Maldacena, H.~Ooguri and Y.~Oz,
``Large $N$ field theories, string theory and gravity'',
Phys.\ Rept.\  {\bf 323} (2000) 183,
hep-th/9905111.}
\lref\twozerorefs{E.~Witten,
``Some comments on string dynamics'',
hep-th/9507121;N.~Seiberg,
``New theories in six dimensions and matrix description of M-theory on  
$T^5$ and $T^5/Z_2$'',
Phys.\ Lett.\  {\bf B408}, 98 (1997).
hep-th/9705221.}
\lref\dualityrefs{C.~Montonen and D.~Olive,
``Magnetic Monopoles As Gauge Particles?'',
Phys.\ Lett.\  {\bf B72}, 117 (1977); A.~Sen,
``Strong - weak coupling duality in four-dimensional string theory'',
Int.\ J.\ Mod.\ Phys.\  {\bf A9}, 3707 (1994).
hep-th/9402002.}
\lref\verlinde{E.~Verlinde,
``On the holographic principle in a radiation dominated universe'',
hep-th/0008140.}
\lref\cardyh{J.~L.~Cardy,
``Operator content and modular properties of higher dimensional conformal 
field theories''. Nucl.\ Phys.\  {\bf B366} (1991) 403.}
\lref\cardytwod{J.~L.~Cardy,
``Operator Content Of Two-Dimensional Conformally Invariant Theories,''
Nucl.\ Phys.\  {\bf B270} (1986) 186.}
\lref\wittenthermo{E.~Witten,
``Anti-de Sitter space, thermal phase transition, and confinement in  
gauge theories'',
Adv.\ Theor.\ Math.\ Phys.\  {\bf 2} (1998) 505, hep-th/9803131.}
\lref\wald{R.~M.~Wald,
``The thermodynamics of black holes'',
gr-qc/9912119.}
\lref\bekbound{J.~D.~Bekenstein,
``A Universal Upper Bound On The Entropy To Energy Ratio For 
Bounded Systems'', Phys.\ Rev.\  {\bf D23} (1981) 287;
``Entropy bounds and black hole remnants'', Phys.\ Rev.\  
{\bf D49} (1994) 1912, gr-qc/9307035.}
\lref\gkp{S.~S.~Gubser, I.~R.~Klebanov and A.~W.~Peet,
``Entropy and Temperature of Black 3-Branes'',
Phys.\ Rev.\  {\bf D54} (1996) 3915,
hep-th/9602135.}
\lref\kimetal{H.~J.~Kim, L.~J.~Romans and P.~van Nieuwenhuizen,
``The Mass Spectrum Of Chiral N=2 D = 10 Supergravity On $S^5$'',
Phys.\ Rev.\  {\bf D32} (1985) 389.}
\lref\anrefs{P.~Candelas and S.~Weinberg,
``Calculation Of Gauge Couplings And Compact Circumferences From 
Selfconsistent Dimensional Reduction'',
Nucl.\ Phys.\  {\bf B237} (1984) 397;
A.~Cappelli and A.~Coste,
``On The Stress Tensor Of Conformal Field Theories In Higher Dimensions'',
Nucl.\ Phys.\  {\bf B314} (1989) 707;
A.~Cappelli and G.~D'Appollonio,
``On the trace anomaly as a measure of degrees of freedom'',
Phys.\ Lett.\  {\bf B487} (2000) 87,
hep-th/0005115.}
\lref\newcon{T.~Appelquist, A.~G.~Cohen and M.~Schmaltz,
``A new constraint on strongly coupled field theories'',
Phys.\ Rev.\  {\bf D60} (1999) 045003,
hep-th/9901109; ``New constraints on chiral gauge theories'',
Phys.\ Lett.\  {\bf B459} (1999) 235,
hep-th/9904172.}

\newsec{Introduction}

Twenty years ago, Bekenstein proposed \bekbound\
that the entropy of a complete physical system in
three spatial dimensions whose total energy is $E$
and which fits in a sphere of radius $R$, is
necessarily bounded from above,
\eqn\bbeekk{S\le 2\pi ER~.}
The motivation for the bound came from studying
the consequences of the generalized second law (GSL)
of thermodynamics (\ie\ thermodynamics in the presence
of black holes). When an entropy-bearing object is
dropped into a black hole, the GSL appears to be
violated unless the infalling object satisfies the
bound \bbeekk. A similar bound should
hold by the same arguments in other spacetime dimensions.

\lref\bekrev{J.D. Bekenstein, 
``On Page's examples challenging the entropy bound'', gr-qc/0006003.}

Over the years, various objections to the bound \bbeekk\
have been raised (see \refs{\wald,\bekrev} for recent
reviews). In addition to the fact that it is not clear
how to define $E$ and $R$ in a general non-spherically
symmetric spacetime, it is now also not clear that the bound
\bbeekk\ is necessary for the validity of the GSL \wald.

An interesting feature of the bound \bbeekk\ is that it does
not involve the gravitational coupling constant $G_N$, and thus
should remain valid in the limit $G_N\to 0$. In other words it
should be a property of the non-gravitational dynamics in a fixed
background spacetime, and can be checked directly in model
systems. Many such checks have been performed \bekrev, but
the situation in some of these cases and in general seems
unclear.

A natural class of theories in which to study the bound
\bbeekk\ is quantum field theories in $D$ dimensions, and
in particular the conformal field theories (CFT's) to
which they flow in the extreme UV and IR limits. The
latter problem was discussed in an interesting recent paper
by E. Verlinde \verlinde, who argued that the entropy of a
general $D$-dimensional CFT on
\eqn\sptime{\IR\times S^{D-1}~,}
is related to the energy via
\eqn\verent{
S = {2\pi R\over D-1}\sqrt{E_C(2E-E_C)}~,
}
where $R$ is the radius of $S^{D-1}$, and $E_{C}$ is the
sub-extensive part of the energy $E$. It is defined through
\eqn\suben{
E_{C} \equiv D E - (D-1) TS~,
}
and is parametrically smaller than the energy for high
temperature, $TR\gg 1$. The evidence supporting \verent\
is twofold:
\item{(1)}
In $D=2$, \verent\ reduces to Cardy's asymptotic
entropy formula \cardytwod
\eqn\enttd{
S = 2\pi\sqrt{{c\over 6}(L_0-{c\over 24})}~,
}
which is valid\foot{More precisely, \enttd\ is valid in
unitary CFT's with $c\gg 1$ and a gap in the spectrum of
scaling dimensions, for $L_0\gg {c\over24}$.} for general
CFT's in $D=2$.
\item{(2)}
According to the AdS/CFT correspondence \adsrev\ a large
class of CFT's are described by a holographic dual.
For such theories, \verent\ is valid in the strong
coupling regime where the dual theory reduces to
supergravity \verlinde, for all energies above the
Hawking-Page phase transition.
\medskip
\noindent
A generalization like \verent\ of the Cardy formula \enttd\
that is universally true would be very interesting, both
because it implies that the relation \bbeekk\ is always
satisfied in the region of validity of \verent, and for
other applications. Since it seems to hold for strongly
coupled CFT's with $AdS$ duals \verlinde, it is natural
to ask whether it also holds in the opposite limit of
vanishing coupling. This question is analyzed in this paper;
unfortunately the answer is negative. We arrive at this
conclusion by computing the partition functions of different CFT's 
and analyzing their high temperature thermodynamics. 
The techniques we use are applicable for general dimension $D$, 
but we study in detail the cases of $D=4$ and $D=6$.

Before turning to the detailed calculations, we summarize
some of the qualitative features that are common to the
different examples we study.
\item{(1)} The high temperature expansion of the free energy 
has the form
\eqn\freeen{
-FR = a_D (2\pi RT)^D + a_{D-2}(2\pi RT)^{D-2}+ \cdots +
a_0 (2\pi RT)^0 +{\cal O}(e^{-(2\pi)^2 RT})~.
}
The perturbative part of the free energy is thus a finite
polynomial {\it with no inverse powers in the temperature}.
The non-perturbative corrections are of a specific form,
namely a power series in $e^{-(2\pi)^2TR}$.
\item{(2)}
The coefficients in the expansion \freeen\ satisfy
\eqn\sumrule{
\sum_{k=0}^{D/2}(-1)^{k}(2k-1)a_{2k}=0~.
}
In two dimensions this gives $a_{2}+a_{0}=0$ --- the
leading term of ${\cal O}((RT)^2)$ is identical (up to
a sign) to the term independent of the temperature. This
is indeed a well-known feature of (unitary) CFT's in $D=2$.
The sum rule \sumrule\ generalizes this to higher dimensions.
The polynomial part of the free energy is characterized by $D/2$
coefficients and one constraint, for a net $D/2-1$ parameters,
whose values in the different examples are given below in 
(4.5) and (4.11).
\item{(3)} 
It is surprisingly useful to study the transformation under exchange
of the radii $\beta/2\pi$ and $R$ of the $S^1$ and the $S^{(D-1)}$,
generalizing a modular transformation in $D=2$. In higher dimensions 
this is generally not a full-fledged symmetry; nevertheless free field
partition 
functions transform in a simple way.
\medskip
\noindent
The paper is organized as follows. In section 2 we compute the
partition functions for various conformal field theories. Section
3 contains an analysis of the high temperature expansions of
certain building blocks of these partition functions. The result
is used in section 4 to find the high temperature expansions
of the free energies of these CFT's. In section 5 we discuss
the implications of our results to entropy bounds.


\newsec{The Partition Functions of Free CFTs}

To study the thermodynamics of CFT on \sptime\
one is interested in the spectrum of the Hamiltonian
on $S^{D-1}$. The coordinate change
$\tau = R{\rm ln}{\rho\over R}$ maps the line element
\eqn\lineelone{
ds^2 = d\tau^2 + R^2 d\Omega^2_{D-1}~,
}
on \sptime\ (with Euclidean time) to
\eqn\lineeltwo{
ds^2 = {R^2\over\rho^2}(d\rho^2 + \rho^2 d\Omega^2_{D-1})~,
}
and makes it manifest that $\IR\times S^{D-1}$ is conformally
equivalent to the Euclidian space $\IR^D$. Under this map the
generator of time translations on \lineelone\ is related to
the generator of scale transformations on \lineeltwo\ as
\eqn\rrho{
R{\partial\over\partial\tau} = \rho {\partial\over\partial\rho}~.
}
The energy $E$ of a state on $\IR\times S^{D-1}$ is therefore
related to the scaling dimension of the corresponding field as
\eqn\erdelta{ER = \Delta~.}
The partition sum
\eqn\partsumz{Z={\rm Tr} e^{-\beta H}}
of the CFT on \sptime\ is obtained by compactifying
Euclidean time $\tau$ \lineelone\ on a circle of
circumference $\beta$, $\tau\sim \tau + \beta$.
Via the correspondence  \erdelta\ it can be reinterpreted
as the generating functional of conformal dimensions on
$\IR^D$,
\eqn\genfndim{Z = \sum_\Delta q^\Delta~,}
with
\eqn\qbr{q=e^{-{\beta\over R}}.}
To compute it one can either enumerate local operators
in the CFT on $\IR^{D}$, or perform the quadratic path
integral directly by analyzing the appropriate wave equation
on $S^1\times S^{D-1}$. In the free theory either approach is
manageable. In the following we carry out each of these
derivations in the simplest case. We note some group theory
which helps to automate the computation for other fields.

\subsec{Four Dimensions}
We begin by considering the conformally coupled scalar in $D=4$. The
equation of motion is
$$
[- \nabla^2  + \xi {\cal R} ]\phi  = 0~,
$$
where the conformal coupling in $D=4$ is $\xi={1\over 6}$ and the
Ricci tensor of the spatial sphere is ${\cal R}={6\over R^2}$.
The kinetic term $-\nabla^2$ on $S^1\times S^3$ contributes the energy 
$-E^2$ and the centrifugal term ${n(n+2)\over R^2}$ where $n$ is the 
integer partial wave number on the $S^3$. Altogether this gives the
spectrum
\eqn\scalaspec{
\Delta = ER = n+1~~~;~~n=0,1,\ldots~.
}
The degeneracy comes exclusively from the wave functions on
the sphere $S^3$. The dimension of the spin
$({n\over 2},{n\over 2})$ representation of
$SO(4)\simeq SU(2)\times SU(2)$ corresponding to the
$n$'th partial wave is $(n+1)^2$.

This type of computation can be repeated for other fields. However,
it is generally awkward to keep track of the couplings to the background
curvature for fields other than the scalar and those details will at any 
rate turn out unimportant; so we will not record them here (see {\it e.g.}
\anrefs\ for further details).

We now turn to the alternative strategy of enumerating free operators 
on the flat background $\IR^4$ \cardyh, considering again the 
conformally coupled scalar.
The field $\phi$ has dimension $\Delta=1$. Its
derivatives $\partial_\mu\phi$ have dimension $\Delta=2$ and
degeneracy $4$ due to the indices $\mu$. At the next level
there are ${5\cdot 4\over 2} =10$ operators $\partial_\mu\partial_\nu\phi$
with $\Delta=3$. But now there is also a constraint $-\nabla^2\phi=0$
from the equation of motion (in flat space) so there is a net 
degeneracy $9$. Proceeding similarly to all orders one recovers 
\scalaspec\ with the expected degeneracy $(n+1)^2$.

It is instructive to carry out this sort of counting for the first few
levels of all the CFTs we consider. However, it is clearly tedious and
not particularly illuminating to work out the combinatorics at arbitrary 
level. It is simpler to infer the complete tower of $SO(4)$ 
representations from the existence of a partial wave expansion. 
This reasoning can be justified as follows. At level $\Delta=n+1$
the field transforms in the $({n\over 2},{n\over 2})$ of $SO(4)$. 
Acting with the derivative $\partial_\mu$ yields
\eqn\groupp{
\left. ({1\over 2},{1\over 2})\otimes
({n\over 2},{n\over 2})\right|_{\rm sym}
= ({n+1\over 2},{n+1\over 2})\oplus ({n-1\over 2},{n-1\over 2})~.
}
The restriction to the symmetric part reflects symmetrization of 
multiple derivatives. Now, the field equation transforms as  
$({n-1\over 2},{n-1\over 2})$ and removes the final term so we are left
with $({n+1\over 2},{n+1\over 2})$, the appropriate representation
content at one level higher. In this way we find a full tower of operators.
This computation generalizes to fields with spin. 

Let us consider the Maxwell field. The field strength $F_{\mu\nu}$
has dimension $\Delta=2$ and there are $6$ components. At the next
level the operators $\partial_\lambda F_{\mu\nu}$ with $\Delta=3$
have $24$ components. The constraints imposed by the $4$ field equations 
$\partial^\mu F_{\mu\nu}=0$ and the $4$ Bianchi identities 
$\partial^\mu {}^*F_{\mu\nu}=0$ yield net degeneracy of $16$. At level 
$\Delta=4$ similar computations give a degeneracy of $40$. These 
degeneracies agree with those of two $SO(4)$ towers of the form 
$({n\over 2},{n+2\over 2})+({n+2\over 2},{n\over 2})$. The 
precise representations were determined from the $SO(4)$ content 
at low level. In the free field limit it is in fact consistent to 
consider the two helicities independently (the self-dual and the anti-self 
dual part of the field).

Repeating the exercise for fermions completes the following table:
\bigskip
\vbox{
$$\vbox{\offinterlineskip
\hrule height 1.1pt
\halign{&\vrule width 1.1pt#
&\strut\quad#\hfil\quad&
\vrule width 1.1pt#
&\strut\quad#\hfil\quad&
\vrule width 1.1pt#
&\strut\quad#\hfil\quad&
\vrule width 1.1pt#
&\strut\quad#\hfil\quad&
\vrule width 1.1pt#\cr
height3pt
&\omit&
&\omit&
&\omit&
&\omit&
\cr
&\hfil field &
&\hfil $\Delta$&
&\hfil degeneracy&
&\hfil $SO(4)$ rep.&
\cr
height3pt
&\omit&
&\omit&
&\omit&
&\omit&
\cr
\noalign{\hrule height 1.1pt}
height3pt
&\omit&
&\omit&
&\omit&
&\omit&
\cr
&\hfil Scalar&
&\hfil $n+1$&
&\hfil $(n+1)^2=1,4,9,\ldots$&
&\hfil $({n\over 2},{n\over 2})$&
\cr
height3pt
&\omit&
&\omit&
&\omit&
&\omit&
\cr
\noalign{\hrule}
height3pt
&\omit&
&\omit&
&\omit&
&\omit&
\cr
&\hfil Weyl fermion&
&\hfil $n+{3\over 2}$&
&\hfil $2(n+1)(n+2)=4,12,24,\ldots$&
&\hfil $2({n\over 2},{n+1\over 2})$&
\cr
\noalign{\hrule}
height3pt
&\omit&
&\omit&
&\omit&
&\omit&
\cr
&\hfil Vector&
&\hfil $n+2$&
&\hfil $2(n+1)(n+3)=6,16,30,\ldots$&
&\hfil $({n\over 2},{n+2\over 2})+h.c.$&
\cr
&\omit&
&\omit&
&\omit&
&\omit&
\cr
}\hrule height 1.1pt
}
$$
}
\centerline{\sl Table 1: spectrum of free fields in four dimensions.
The range of $n=0,1,\ldots$.}
\centerline{\sl A definite chirality was chosen for the Weyl fermion.}
\centerline{\sl Both chiralities were included for the vector field 
(Maxwell case).}
\def\d4states{Table 1}
\bigskip

Alternatively the results can be summarized concisely in terms of the
partition functions \genfndim:
\eqn\zfourd{\eqalign{
Z_S^{(4)} =& \prod_{n=0}^\infty \left(
{1\over 1-q^{n+1}}\right)^{(n+1)^2}~,\cr
Z_W^{(4)}=& \prod_{n=0}^\infty \left(
1+q^{n+{1\over 2}}\right)^{2n(n+1)}~,\cr
Z_V =& \prod_{n=0}^\infty \left(
{1\over 1-q^{n+1}}\right)^{2n(n+2)}~.
}}
Note that for the fermion and the vector the range of the index $n$ 
was shifted compared with the table without changing the partition 
function.

$N=4$ SYM theory in $D=4$ with gauge group $U(1)$ has the field 
content: $6$ scalars, $4$ Weyl fermions, and one vector field. It 
therefore has the partition function
$$
Z_{N=4} = (Z_S^{(4)})^6~(Z_W^{(4)})^4~Z_V~.
$$
Different gauge groups (with more gauge fields) can be incorporated in 
the free field limit by simply taking the appropriate power of the 
entire partition function.

\lref\slansky{R. Slansky, Phys. Rept. {\bf 79} (1981) 1.}

\subsec{Six Dimensions}

The analysis of the six dimensional case is quite similar to the 
discussion for $D=4$ so we shall be brief. Let us consider the
self-dual tensor which may be less familiar. It is described by the
field strength $H_{\mu\nu\rho}$ with conformal dimension $\Delta=3$. 
Taking the antisymmetry in each index into account this gives
$6\cdot 5\cdot 4/3!=20$ components in six dimensions. The duality 
condition ${}^*H=H$ plays the role of the field equation. It removes
half the components for a net degeneracy of $10$. At the next level
there are the fields $\partial_\lambda H_{\mu\nu\rho}$ with $\Delta=4$. 
There are $6\cdot 10$ such components after the duality
condition on $H$ has been taken into account. 
The trace $\partial^\mu H_{\mu\nu\rho}=0$
gives further $15$ conditions for a net degeneracy of $45$. Note that
the equation $\partial^\mu {~}^*H_{\mu\nu\rho}=0$ gives nothing new. It is
now clear that the tensor field generate a tower with $SO(6)$ quantum
numbers $[2,n,0]$ with $n=0,1,\ldots$ (see \eg\ \slansky\ for further
details on these representations).
The conjugate tower $[0,n,2]$ corresponds to a field with the
opposite duality condition.

The spectrum of the free fields in six dimensions is
\bigskip
\vbox{
$$\vbox{\offinterlineskip
\hrule height 1.1pt
\halign{&\vrule width 1.1pt#
&\strut\quad#\hfil\quad&
\vrule width 1.1pt#
&\strut\quad#\hfil\quad&
\vrule width 1.1pt#
&\strut\quad#\hfil\quad&
\vrule width 1.1pt#
&\strut\quad#\hfil\quad&
\vrule width 1.1pt#\cr
height3pt
&\omit&
&\omit&
&\omit&
&\omit&
\cr
&\hfil field &
&\hfil $\Delta$&
&\hfil degeneracy&
&\hfil $SO(6)$ rep.&
\cr
height3pt
&\omit&
&\omit&
&\omit&
&\omit&
\cr
\noalign{\hrule height 1.1pt}
height3pt
&\omit&
&\omit&
&\omit&
&\omit&
\cr
&\hfil Scalar&
&\hfil $n+2$&
&\hfil ${(n+1)(n+2)^2(n+3)\over 12}=1,6,20\ldots$&
&\hfil $[0,n,0]$&
\cr
height3pt
&\omit&
&\omit&
&\omit&
&\omit&
\cr
\noalign{\hrule}
height3pt
&\omit&
&\omit&
&\omit&
&\omit&
\cr
&\hfil Weyl fermion&
&\hfil $n+{5\over 2}$&
&\hfil ${(n+1)(n+2)(n+3)(n+4)\over 3}=8,40,120,\ldots$&
&\hfil $[1,n,0]$&
\cr
\noalign{\hrule}
height3pt
&\omit&
&\omit&
&\omit&
&\omit&
\cr
&\hfil Tensor&
&\hfil $n+3$&
&\hfil ${(n+1)(n+2)(n+4)(n+5)\over 4}=10,45,126,\ldots$&
&\hfil $[2,n,0]$&
\cr
&\omit&
&\omit&
&\omit&
&\omit&
\cr
}\hrule height 1.1pt
}
$$
}
\centerline{\sl Table 2: spectrum of free fields in six dimensions. The 
range of $n=0,1,\ldots$.}
\centerline{\sl A definite chirality was chosen for the Weyl fermion and 
the Tensor.}
\def\d6states{Table 2}
\bigskip

The corresponding partition functions are
\eqn\zsixd{\eqalign{
Z_S^{(6)} =& \prod_{n=0}^\infty \left( 
{1\over 1-q^{n+1}}\right)^{n(n+1)^2(n+2)/12}~,\cr
Z_W^{(6)}=& \prod_{n=0}^\infty 
\left( 1+q^{n+{1\over 2}}\right)^{(n-1)n(n+1)(n+2)/3}~,\cr
Z_T =& \prod_{n=0}^\infty
\left({1\over 1-q^{n+1}}\right)^{(n-1)n(n+2)(n+3)/4}~.}}

The $(2,0)$ field theory in $D=6$ has the field content: $5$ scalars, 
$2$ Weyl fermions, and one tensor field. It therefore has partition function
$$
Z_{(2,0)} = (Z_S^{(6)})^5~(Z_W^{(6)})^2~Z_T~.
$$
Again we can consider a theory with $N$ independent species,
described by taking the entire partition function to the appropriate
power. This is relevant for describing the $(2,0)$ CFT corresponding
to $N$ fivebranes at a generic point on the moduli space of vacua
(where the fivebranes are separated). Coincident fivebranes are
described by a strongly interacting CFT to which the above discussion
does not apply.

\newsec{Analysis of Partition Functions}

To construct the high temperature thermodynamics of the
partition sums \zfourd, \zsixd, we start in this section
by analyzing the properties of certain basic partition
functions. In the following section the results are applied
to study the specific partition functions found in section 2.

\subsec{The Bosons}
The partition function
\eqn\zbdef{
Z_B^{(d)} = \prod_{n=0}^\infty \left(
{1\over 1-q^{n+1}}\right)^{(n+1)^{d-2}}~,
}
can be interpreted as a simple model for a bosonic field in $D=d$ 
dimensions. We assume that the parameter $d$ is even and use 
the notation $q=e^{-2\pi\delta}=e^{-\beta/R}$ {\it i.e.}
$\delta =(2\pi RT)^{-1}$.

In the following we generalize a computation by Cardy \cardyh\ .
First, take the logarithm of the partition sum and expand
$$
-\delta^{d/2}{\partial\over\partial\delta} {\rm ln} Z_B^{(d)}
= 2\pi\sum_{n=0}^{\infty} (n+1)^{d-1}\delta^{d/2}\sum_{k=1}^{\infty} 
e^{-2\pi k(n+1)\delta}~.
$$
Then use the representation
$$
e^{-x} = {1\over 2\pi i}\int_C x^{-(s+d/2)}\Gamma(s+{d\over 2})ds~,
$$
where the contour $C$ is along the imaginary axis, with the real part 
of $s$ large ($\Re s>d/2$). Finally arrive at the Mellin transform
\eqn\zdtwo{
-\delta^{d/2}{\partial\over\partial\delta} {\rm ln} Z_B^{(d)}
= {1\over 2\pi i}\int_C \delta^{-s}G_B^{(d)}(s)ds~,
}
where
$$
G_B^{(d)}(s) = (2\pi)^{-s+1-k/2}~\Gamma(s+{d\over 2})~\zeta(s+{d\over 2})
~\zeta(s+1-{d\over 2})~.
$$
The function $G_B^{(d)}$ is meromorphic with poles
\eqn\pospole{
G_B^{(d)}(s) \sim 2\pi~{|B_d|\over 2d}~{1\over s-{d\over 2}}~~~;~~ 
s\sim {d\over 2}~,
}
and
\eqn\negpole{
G_B^{(d)}(s) \sim 2\pi~{B_d\over 2d}~{1\over s+{d\over 2}}~~~;~~ 
s\sim -{d\over 2}~.
}
We used 
$$
\zeta(2n) = {(2\pi)^{2n}\over 2(2n)!} |B_{2n}|~~~;~~
\zeta(1-2n) = - {B_{2n}\over 2n}~,
$$
for $n\in\ZZ_{+}$, and also $\zeta(0)=-{1\over 2}$. 
Here $B_d$ are the Bernoulli numbers $B_2={1\over 6}$, $B_4=-{1\over 30}$,
$B_6={1\over 42}$ {\it etc}. For $d=2$ there is an additional pole
$$
G_B^{(2)}(s) \sim  - {1\over 2s}~~~;~~ 
s\sim 0~.
$$
Integration around the poles in \zdtwo\ yields an expression for
the partition functions at small $\delta$:
\eqn\powlaw{
{\rm ln}Z_B^{(d)}\simeq 2\pi~{|B_d|\over 2d}~
\left({1\over d-1}\delta^{-d+1}+(-1)^{d/2}\delta\right) + {1\over 2}{\rm ln}
\delta~\delta_{d,2}~.
}
For reference we write out the first few partition functions
\eqn\fewz{\eqalign{
{\rm ln}Z_B^{(2)}\simeq &~2\pi~{1\over 24}~
(\delta^{-1}-\delta) + {1\over 2}{\rm ln}\delta~,\cr
{\rm ln}Z_B^{(4)}\simeq &~2\pi~{1\over 240}~
({1\over 3}\delta^{-3}+\delta) ~,\cr
{\rm ln}Z_B^{(6)}\simeq &~2\pi~{1\over 504}~
({1\over 5}\delta^{-5}-\delta)~. 
}}
In the following we will use modular invariance to determine that the 
corrections to each of these formulae are exponentially small for large 
$RT$, of ${\cal O}(e^{-(2\pi)^2 RT})$. 

As a preliminary result note that the identity
$$
\zeta(s) = 2^s \pi^{s+1} ~{\rm sin}{\pi s\over 2} 
~\Gamma(1-s)~\zeta(1-s)~,
$$
gives the symmetry property
\eqn\gsymm{
G_B^{(d)}(-s) = (-)^{d/2}G_B^{(d)}(s)~.
}
Recalling $|B_d|=(-1)^{d/2}B_d$ for even $d$ this is consistent
with the obvious symmetry between the poles given in \pospole\
and \negpole. Before proceeding we introduce the notation
$$
I_B^{(d)}(\delta) = - \delta^{d/2} {\partial\over\partial\delta}
{\rm ln}\left[q^{-\gamma^{B}_d}Z_B^{(d)}\right]~,
$$
where
\eqn\gamshift{
\gamma_d^{B} = {B_d\over 2d}~,
}
generalizes the shift $c/24=1/24$ familiar in the CFT of a free
boson in $D=2$. Now, deform the contour $C$ in \zdtwo\ from large
positive $\Re s$ to large negative $\Re s$ by integrating over the
poles. Then use the symmetry \gsymm\ to return the contour to large
positive $\Re s$. The result can be written
\eqn\imodinv{
I_B^{(d)}({1\over\delta})=(-1)^{d/2}~I_B^{(d)}(\delta)~.
}
The role of the shift \gamshift\ is to take the poles correctly into 
account. For brevity we have omitted the zero-mode for $d=2$.

The modular relation \imodinv\ gives an alternative derivation of the 
power law terms \powlaw\ (and therefore \fewz\ ). Simply use \imodinv\ to 
relate the behavior at small $\delta$ to that of large $\delta$; and 
then note that at large $\delta$ \zbdef\ makes it manifest 
that the partition function is a power series in 
$e^{-2\pi\delta}=e^{-(2\pi)^{2}RT}$.

\subsec{The Fermions}
The considerations above can be repeated for fermions. 
The basic partition function is
$$
Z_F^{(d)} = \prod_{n=0}^\infty \left(1+
q^{n+{1\over 2}}\right)^{(n+{1\over 2})^{d-2}}~.
$$
Carrying out the steps \zbdef\ through \zdtwo\
we find the Mellin transform
$$
-\delta^{d/2}{\partial\over\partial\delta} {\rm ln} Z_F^{(d)}
= {1\over 2\pi i}\int_C \delta^{-s}G_F^{(d)}(s)ds~,
$$
where
$$
G_F^{(d)}(s) = G_B^{(d)}(s)\left(1-{1\over 2^{s-1+{d\over 2}}}\right)
\left( 2^{s-{d\over 2}+1} -1 \right)~.
$$
The pole of $G_B^{(d)}$ at $s=0$ for $d=2$ is cancelled by
the first term in brackets. This reflects the absence of
zero-modes for (NS) fermions. The only poles are therefore
$$
G_F^{(d)}(s) \sim 2\pi~\left( 1 - {1\over 2^{d-1}}\right)
{|B_d|\over 2d}~{1\over s-{d\over 2}}~~~;~~ s\sim {d\over 2}~,
$$  
and its mirror image under the symmetry 
$$
G_F^{(d)}(-s) = (-)^{d/2}G_F^{(d)}(s)~.
$$
This symmetry of $G^{(d)}_F$ is the key step in 
showing that modular symmetry is maintained for fermions. In other words
\eqn\imodfinv{
I_F^{(d)}({1\over\delta})=(-1)^{d/2}~I_F^{(d)}(\delta)~,
} 
where now 
$$
I_F^{(d)}(\delta) = - \delta^{d/2} {\partial\over\partial\delta}
{\rm ln}\left[q^{-\gamma^{F}_d}Z_F^{(d)}\right]~.
$$
We introduced
\eqn\gamfshift{
\gamma_d^{F} = (1-{1\over 2^{d-1}})\gamma_d^{B} =
(1-{1\over 2^{d-1}}){B_d\over 2d}~,
}
generalizing the shift $c/24=1/48$ familiar from the CFT of a free (NS) 
fermion in $d=2$.

The leading (power series) behavior of the partition functions is related
to that of the bosons in a simple way
$$
{\rm ln} Z^{(d)}_F \simeq \left( 1 - {1\over 2^{d-1}}\right)
{\rm ln} Z^{(d)}_B~.
$$
This is well-known for the thermodynamic behavior at large
temperature; the modular invariance implies
that it also holds for the Casimir corrections.

There are more general boundary conditions that could be
analyzed similarly ({\it e.g.} generalizing R fermions to
$d$ dimensions). These functions do not appear in the
examples considered in this paper.

\newsec{Statistical Mechanics of Free Fields}

In this section we use the results of section 3 to analyze
the partition functions of physical interest.

\subsec{Four Dimensions}

The simplest way to analyze the partition functions
\zfourd\ for the various fields is to decompose them
into the basic ones considered in detail in the previous
sections. The result is
\eqn\decfour{\eqalign{
Z^{(4)}_S =& Z_B^{(4)}~, \cr
Z^{(4)}_W =& (Z_F^{(4)})^2~(Z_F^{(2)})^{-{1\over 2}}~,\cr
Z_V =& (Z^{(4)}_B)^2~(Z_B^{(2)})^{-2}~.
}}
The qualitative features of the partition functions therefore
follow from those of the basic ones analyzed in section 3.
Note however that the ``two dimensional'' partition
functions also enter the results in four dimensions.
Since the precise form of the modular symmetry depends
on the index $d$ it follows that the full partition
functions in $D=4$ do {\it not} satisfy any modular
symmetry, except for the boson (the case considered by
Cardy \cardyh). This is not a problem because there is no
principle requiring this kind of symmetry in dimensions larger
than $D=2$ (as far as we are aware).
Thus it would have been surprising if the partition functions 
did satisfy a modular symmetry. For our purposes the important point
is that the modular symmetry {\it of the constituent partition functions} 
suffices to analyze the high temperature behavior of the partition
sums \decfour. 
For example it ensures that the free energies are simple power
series with no inverse powers of temperature, up to non-perturbative
corrections.

The ${\cal N}=4$ SYM has the partition function
\eqn\znfour{
Z_{N=4} = (Z_B^{(4)})^8 ~(Z_B^{(2)})^{-2}~(Z_F^{(4)})^8 ~(Z_F^{(2)})^{-2}~.
}
For the purpose of analyzing the corresponding thermodynamics we can
relate the fermionic partition function to that of the boson and find
\eqn\lnzfour{\eqalign{
{\rm ln}Z_{N=4}\simeq & ~15 ~{\rm ln} Z^{(4)}_B - 3 ~{\rm ln} Z^{(2)}_B\cr
\simeq & ~2\pi\left[ {1\over 16}({1\over 3}\delta^{-3}+\delta)
-{1\over 8}(\delta^{-1}-\delta)\right]~,
}}
up to exponential corrections. More generally, for a theory
with $n_S$ scalars, $n_F$ Weyl fermions, and $n_V$ Maxwell
fields the polynomial part of the partition function is
\eqn\frefourd{\eqalign{
-FR =& ~{1\over 2\pi\delta}~ {\rm ln} Z = 
~a_4 \delta^{-4} + a_2 \delta^{-2} + a_0 \cr
=& a_{4}(2\pi RT)^{4}+a_{2}(2\pi RT)^{2}+a_{0}(2\pi RT)^{0}~,
}}
where
\eqn\afourc{\eqalign{
a_4 =& {1\over 720}\left( n_S + 2n_V + {7\over 4}n_F\right)~,\cr
a_2 =& -{1\over 24}\left( 2n_V + {1\over 4}n_F\right)~,\cr
a_0 =& {1\over 240}\left( n_S + 22n_V + {17\over 4}n_F\right)~.
}}
These three coefficients satisfy the constraint
\eqn\sumfour{
3a_4 - a_2 - a_0 = 0~.
}
This relation is no surprise because it is obviously satisfied 
by the ``constituent'' partition functions \fewz. It shows that the
polynomial part of the free energy in $D=4$ in general depends 
on {\it two} independent parameters.

The generalization \sumrule\ of \sumfour\ to arbitrary dimensions 
follows similarly from the formulae in section 3. 

\subsec{Six Dimensions}

The six dimensional partition functions \zsixd\ can be similarly 
decomposed into the basic ones. The result is
\eqn\decsix{\eqalign{
Z^{(6)}_S =& (Z_B^{(6)})^{1\over 12}  (Z_B^{(4)})^{-{1\over 12}}~, \cr
Z^{(6)}_W =& (Z_F^{(6)})^{1\over 3}(Z_F^{(4)})^{-{5\over 6}}
(Z_F^{(2)})^{3\over 16}~, \cr
Z_T =& (Z^{(6)}_B)^{1\over 4}~(Z_B^{(4)})^{-{5\over 4}}Z_B^{(2)}~.
}}
Note that here all fields, including the scalar, receive
contributions from basic partition functions of the ``wrong''
dimension.

Combining the results for the field content of the $(2,0)$ theory gives
\eqn\ztwozero{
Z_{(2,0)}=
(Z_B^{(6)})^{2\over 3}  ~(Z_B^{(4)})^{-{5\over 3}} ~Z_B^{(2)}~
(Z_F^{(6)})^{2\over 3}  ~(Z_F^{(4)})^{-{5\over 3}} ~(Z_F^{(2)})^{3\over 
8}~.
}
The leading behavior for small $\delta$ becomes
\eqn\apxlogz{\eqalign{
{\rm ln}Z_{(2,0)}\simeq &
~{2\over 3}~{63\over 32}{\rm ln}Z_B^{(6)}
-{5\over 3}~{15\over 8}{\rm ln}Z_B^{(4)}
+(1+{3\over 16}){\rm ln}Z_B^{(2)}\cr
\simeq & ~2\pi~{1\over 12\cdot 32}
\left[ ({1\over 5}\delta^{-5}-\delta)
-5({1\over 3}\delta^{-3} + \delta) + 19(\delta^{-1}-\delta)\right]~,
}}
up to exponential corrections.

The partition functions \znfour\ and \ztwozero\ for the maximally 
symmetric theories share an interesting property. In each case
$Z_B^{(d)}$ and $Z_F^{(d)}$ appear to the same power,
as they must due to supersymmetry. The same is true in both cases
for the first subleading terms $Z_B^{(d-2)}$ and $Z_F^{(d-2)}$.
It is possible that this is also guaranteed by the large
supersymmetry, but we have not analyzed this.

For a general theory with $n_S$ scalars, $n_F$ Weyl fermions,
and $n_T$ tensor fields the polynomial part of the partition
function is
\eqn\fresixd{
-FR = a_6 \delta^{-6} + a_4 \delta^{-4} + a_2 \delta^{-2} + a_0 
}
where
\eqn\asixc{\eqalign{
a_6 =& {1\over 2520}\left({1\over 12}n_{S} + {1\over 4}n_{T}+ {31\over 
192}n_{F}\right)~,\cr
a_4 =& -{1\over 720}\left( {1\over 12}n_S + {5\over 4}n_T + 
{35\over 96}n_F\right)~,\cr
a_2 =& {1\over 24} \left( n_T + {3\over 64}n_F\right)~,\cr
a_0 =& -{1\over 4032}\left( {31\over 15}n_S + 191n_T + 
{367\over 24}n_F\right)~.
}}
These four coefficients satisfy the constraint \sumrule
$$
5a_6 - 3a_4 + a_2 + a_0 = 0~.
$$
Thus the polynomial part of the free energy in $D=6$ generally depends 
on {\it three} independent parameters.

\newsec{Entropy Formulae and Bounds}

In this section we apply our study of free CFT's to discuss the entropy
formula \verent\ recently proposed by E. Verlinde \verlinde, and the
Bekenstein bound \bbeekk.

\subsec{An asymptotic entropy formula?}

Starting from the polynomial part of the free energy \freeen\ it
is straightforward to work out other thermodynamic quantitites,
{\it e.g.} the sub-extensive part of the energy \suben\
$$
E_C R = -\left( 2 a_{D-2}\delta^{-D+2} + 4 a_{D-4}\delta^{-D+4}
+\cdots +  Da_{0}\delta^{0} \right)~.
$$
In these manipulations we assume the thermodynamic limit where
the entropy $S=\beta(E-F)$ is found by a saddle point
approximation. An improved treatment would give logarithmic and higher 
corrections that are unimportant at high temperature.

The results of sections 2-4 imply that:
\item{(1)} 
At large temperature the leading sub-extensive energy is 
governed by $a_{D-2}$. For general matter content \afourc\ and \asixc\
give $a_{D-2}\leq 0$. We therefore find $E_{C}\geq 0$ to the leading order. 
This is encouraging because a negative value of $E_{C}$ renders \verent\ 
meaningless. However, there is a specific case where the inequality 
is saturated and $a_{D-2}=0$. This is the conformally coupled scalar 
field in $D=4$. Moreover, here the next order is $a_0>0$ and so $E_C<0$. 
It is clear that some essential modification is needed in this situation.

\item{(2)} 
Leaving aside the conformally coupled scalar in $D=4$ we have 
$a_{D-2}<0$ and so $E_{C}\sim (TR)^{D-2}$ for large $TR$. Therefore
the functional dependence on $TR$ agrees on the two sides 
of \verent\ for large $TR$. The existence of a component of the energy 
with $E_{C} \sim (TR)^{D-2}$ is a key point of the reasoning in \verlinde\ .
However, although the {\it scaling} works correctly for large $TR$ the 
{\it coefficients} do not in general match. For large $TR$, the
leading term in \verent\ gives the relation
\eqn\acond{
a_{D-2} = - {D^2 (D-1)\over 4}~a_D~.
}
Remarkably this is satisfied for theories with a holographic dual 
\refs{\verlinde,\wittenthermo}. However, for the theories considered in 
this paper there is in general no relation between $a_{D-2}$ and $a_{D}$,
except for $D=2$ where $a_0=-a_2$ in agreement with \acond. 
It is therefore not sufficient to modify the overall numerical
coefficient of the r.h.s. of \verent, because
the change would have to depend on the matter content of the theory.

\item{(3)} Finally, let us note that although the two sides of \verent\ 
generically have identical scaling behavior for large $TR$ the full
functional form is in general different once the subleading terms 
are taken into account.

\bigskip
\noindent
In $D=4$, \acond\ can be compared with our result
$$
-{a_2\over a_4} = 30 ~ 
{ 2n_V + {1\over 4}n_F\over n_S + 2n_V + {7\over 4}n_F}~.
$$
As we vary the matter content, this expression covers the
{\it bounded} range
$$
0\leq \left|{a_2\over a_4}\right| \leq 30 ~.
$$
The bounds are saturated for pure scalar and vector theories,
respectively. The corresponding range for $D=6$
$$
{7\over 2}\leq \left|{a_4\over a_6}\right| \leq {35\over 2}~,
$$
is saturated for scalar and tensor fields, respectively. Now let us
assume that \acond\ is really a prediction for the strongly coupled 
theory. Then the bounds above are consistent with the notion that both 
the extensive $a_D$ {\it and} the sub-extensive $a_{D-2}$ coefficients
flow modestly between weak and strong coupling. The physics of these 
coefficients may therefore be analogous to that 
underlying the (in)famous $3/4$ factor for the entropy \gkp.

As an example, for ${\cal N}=4$ SYM in $D=4$,
at {\it strong} coupling the formula \verent\ is valid and
in particular $a_2 = - 12 a_4$ (according to \acond). At {\it weak}
coupling we find $a_2 = -6 a_4$. The leading coefficient in the 
thermodynamics depends on the coupling so that \gkp\
\eqn\threeq{
a_4 (g^2N=\infty) = {3\over 4}a_4 (g^2N=0)~.
}
The corresponding relation for the leading sub-extensive part 
$$
a_2 (g^2N=\infty) = {3\over 2}a_2 (g^2N=0)~.
$$
When cast in this light the numerical differences between strong and
weak coupling are less significant. Indeed, it is perhaps surprising 
that $a_{2}$ (as $a_{4}$) is so similar in the two regimes, having 
in particular the same overall dependence on the number of species 
($N^{2}$ for $SU(N)$ theory).

These comments were for the leading terms; the full functional 
dependence on $RT$ depends dramatically on the coupling. At strong 
coupling the high temperature expansion of the free energy includes 
an {\it infinite} 
series in $1/RT$ \refs{\wittenthermo,\verlinde}. As the coupling $g^2N$ 
goes to zero, the coefficients of all the terms that go like negative 
powers of $RT$ go to zero and one recovers the polynomial \freeen. 
In particular, the relation \verent\ is far from being satisfied at 
small $g^2N$.

\subsec{The role of finite size effects in violating Bekenstein's
bound}

In the high energy limit $ER\gg 1$ the entropy of a CFT on \sptime\
is dominated by the leading term in \freeen. It can be written as
\eqn\therms{S\sim a_{D}^{1\over D} (ER)^{D-1\over D}.}
The Bekenstein bound \bbeekk\ is not very stringent at
high energies, since the actual entropy \therms\ increases slower
with energy than does the bound. Now, the numerical
coefficient $a_{D}$ is essentially the central charge $c$,
or the number of degrees of freedom of the system (see {\it e.g.} 
\newcon\ for a recent discussion).
In particular, it is proportional to the number of fields in the 
non-interacting theory (see\afourc, \asixc). By comparing \bbeekk\
and \therms\ we see that the bound appears to break down for
\eqn\erc{ER\leq a_D~.}
The fact that potential violations of the bound are associated
with low energies is well known \bekrev. It is usually phrased
as the statement that the bound appears to be violated at low
temperatures. What is interesting in \erc\ is that the dependence
on $a_D$ is precisely such that the relevant energy is the same
as the energy at which finite size effects become important.
Equivalently, the temperature at which the bound is violated is
$T\sim 1/R$, independently of $a_D$. This leads to a number of
consequences:
\item{(1)} In order to determine whether or not the bound \bbeekk\
is in fact violated in the regime \erc\ one has to consider subleading
corrections to the asymptotic formula \therms\ which correspond
to sub-extensive terms in the free energy. This has been carried out
in detail in this paper for weakly coupled CFT's.
\item{(2)} The corrections in question are not just the powerlike
corrections in \freeen, but also the exponential ones. Consider
for example the case of two dimensional CFT. Cardy's formula \enttd\
which is obtained by considering the free energy \freeen\ and ignoring
the exponential corrections appears to predict that the entropy is
bounded from above $S\le 2\pi L_0$, but when the bound is close to
being saturated, the temperature $T$ is of order $1/R$. For such
temperatures the exponential corrections in \freeen\ are important.
Similarly in higher dimensions the exponential terms are important
when the bound \bbeekk\ is close to being violated at $RT\sim 1$.
In fact, the notion of temperature is not really well defined in this
regime, and it is much better to work in the microcanonical ensemble
and count states.
\item{(3)} In the free field examples considered in this paper it
is in fact straightforward to violate the Bekenstein bound
\erc. The reason is that if one decreases the energy
to $ER\sim 1$, the explicit counting shows that the
entropy goes like $S\sim \log a_D$ and the bound is badly violated
for a large number of species $a_D>>1$ \wald. The precise energy
at which the bound is first violated depends on the precise definition
of the bound \bbeekk\ (which as mentioned in the introduction
is unclear), but with any reasonable definition it seems natural
that the violation will occur around the energy \erc. It should
be emphasized that while we have not checked this assertion, it
is possible to do so by using the exact formulae for the partition
sums presented in section 2, by studying the $n$ dependence of
the coefficients of $q^n$.
\item{(4)} One might be confused at this point how the bound
\bbeekk\ could be satisfied in the strong coupling region, where
the $AdS$ calculation of \refs{\wittenthermo,\verlinde} seems
to imply that it is. An example from two dimensional CFT might
be useful in this regard. Consider a sigma model with target space
$T^{4N}$ as a toy model of a ``weakly coupled'' CFT in the higher
dimensional setup, and CFT on $T^{4N}/S_N$ as an example of a
``strongly coupled'' CFT. In the $T^{4N}$ case, the free energy
is just $N$ times that of CFT on $T^4$ and the arguments of
point (3) above apply; one expects the bound to be violated
at $L_0$ of order $N$. For $T^{4N}/S_N$ there are very few states
with $L_0$ of order one, and the arguments of point (3) do not
imply that the bound must be violated. In fact according to
\refs{\wittenthermo,\verlinde} it is not violated in this case.
\medskip
\noindent
Regardless of what one finds in any particular model, it seems
that the bound \bbeekk\ is unlikely to apply in general in the
regime \erc\ since while the high energy behavior \therms\ is
universal, the experience in CFT is that the growth in the
number of states for low and intermediate energies is quite
model dependent. It is thus natural to propose that the bound
\bbeekk\ only applies in the thermodynamic limit, when
finite size effects are negligible. Such a modification would
not be acceptable if the bound \bbeekk\ was needed for the
validity of the GSL of thermodynamics, but if it is not needed
(as argued \eg\ in \wald) we see nothing wrong with it.

Finally, in \bekrev\ it was argued that zero point energy
might be crucial in restoring the bound \bbeekk\ even when
it naively seems to break down. In our context this does
not seem to be the case. It is true that in computing the
free energy we have assigned zero energy to the vacuum on
\sptime. For free fields, there is no mystery in the zero
point energy -- it is simply the contribution of the free fields
to the cosmological constant. For example, it cancels between
bosons and fermions in all the supersymmetric cases mentioned
above. It should also be noted, that the $AdS$ calculations
that establish the bound for strongly coupled CFT's
\refs{\wittenthermo,\verlinde}, assign zero energy to the
vacuum on $S^{D-1}$.

\bigskip\medskip\noindent 
{\bf Acknowledgements:} 
We thank G. Moore, R. Myers, E. Verlinde, and R. Wald for discussions.
This work was supported in part by DOE grant
DE-FG0290ER-40560. FL was supported in part by a Robert R. McCormick 
fellowship.
 
\listrefs 
\end